\documentclass[fleqn,10pt]{wlscirep}
\usepackage[utf8]{inputenc}
\usepackage[T1]{fontenc}

\title{A Multiscale Model for El Ni\~no Complexity}

\author[1]{Nan Chen}
\author[2,3,4,*]{Xianghui Fang}
\author[5]{Jin-Yi Yu}

\affil[1]{Department of Mathematics, University of Wisconsin-Madison, Madison, WI, USA}
\affil[2]{Department of Atmospheric and Oceanic Sciences and Institute of Atmospheric Sciences, Fudan University, 220 Handan Rd., Shanghai 200433, China}
\affil[3]{Innovation Center of Ocean and Atmosphere System, Zhuhai Fudan Innovation Research Institute, Zhuhai 518057, China}
\affil[4]{CMA-FDU Joint Laboratory of Marine Meteorology, Shanghai 200438, China}
\affil[5]{Department of Earth System Science, University of California, Irvine, CA, USA}
\affil[*]{fangxh@fudan.edu.cn}

\begin{abstract}
El Ni\~no–Southern Oscillation (ENSO) exhibits diverse characteristics in  spatial pattern, peak intensity, and temporal evolution. Here we develop a three-region multiscale stochastic model to show that the observed ENSO complexity can be explained by combining intraseasonal, interannual, and decadal processes. The model starts with a deterministic three-region system for the interannual variabilities. Then two stochastic processes of the intraseasonal and decadal variation are incorporated. The model can reproduce not only the general properties of the observed ENSO events, but also the  complexity in patterns (e.g., Central Pacific vs. Eastern Pacific events), intensity (e.g., 10-20 year reoccurrence of extreme El Ni\~nos), and temporal evolution (e.g., more multi-year La Ni\~nas than multi-year El Ni\~nos). While conventional conceptual models were typically used to understand the dynamics behind the common properties of ENSO, this model offers a powerful tool to understand and predict ENSO complexity that challenges our understanding of the 21st-century ENSO. 
\end{abstract}
\begin{document}

\flushbottom
\maketitle
%
%
\thispagestyle{empty}


\section*{Introduction}
As one of the most striking interannual climate variations in the world, El Ni\~no–Southern Oscillation (ENSO) manifests as a basin scale air–sea interaction phenomenon characterized by sea surface temperature (SST) anomalies in the equatorial central to eastern Pacific \cite{rasmusson1982variations}. Although evolving in the equatorial Pacific region, ENSO can affect climate, ecosystems, and economies around the world through atmospheric pathways \cite{ropelewski1987global,klein1999remote}. In the classical viewpoint, ENSO was often regarded as a phenomenon with cyclical attributes \cite{jin1997equatorial}, in which the positive and negative phases are El Ni\~no and La Ni\~na, respectively. ENSO is known to show a significant diversity and irregularity \cite{capotondi2015understanding,timmermann2018nino}. Particularly, many studies have suggested that there are at least two types of ENSO \cite{fu1985two,larkin2005global,ashok2007nino,yu2007decadal,kao2009contrasting}. Based on the features during their mature phase, they were named as the eastern Pacific (EP) and the central Pacific (CP) types when the largest SST anomaly is located near the coast of the South America and the dateline region, respectively \cite{yu2007decadal,kao2009contrasting}. The shift of the main heating location has significant impacts on the air–sea coupling processes in the tropical Pacific, which is the way ENSO affecting the global climate, and brings serious challenges to ENSO predictions \cite{barnston2012skill,zheng2014asymmetry,sohn2016strength}. Thus, since the concept of CP El Ni\~no emerged, understanding the differences in the patterns, strengths, evolution processes, physical mechanisms, and global influences between the two types of ENSO has attracted great attention.

By driving strong anomalous eastward surface currents and exciting downwelling equatorial Kelvin waves, the westerly wind bursts (WWBs), an intraseasonal atmospheric variability, play an important role in the development of El Ni\~no events \cite{mcphaden2004evolution,fedorov2015impact}. Some studies argued that ENSO is likely a result of the interplay between a self-sustaining cyclic oscillation dictated by deterministic processes and WWBs that are partially modulated by ENSO itself \cite{chen2015strong}, in which the former provides a basic dynamical framework, and the latter induces the different flavors of ENSO \cite{lian2018linkage}. Particularly, strong and congregated WWBs are crucial for producing extreme El Ni\~nos \cite{chen2015strong,levine2016extreme}. It has been shown that the stalled El Ni\~no in the winter of 2014 and the “unexpected” extreme El Ni\~no in 2015 could be attributed to the lack and occurrence of WWBs in the spring and summer of 2014 and 2015, respectively \cite{mcphaden2015playing,levine2016extreme,xie2020unusual}. The important role of both WWBs and easterly wind bursts (EWBs) in inducing this delayed extreme El Ni\~no was also highlighted \cite{hu2016exceptionally, hu2017extreme}. Therefore, the stochastic nature of wind bursts can help to explain the irregularity of ENSO events \cite{puy2016modulation}.

The evolution of ENSO is also significantly modulated by physical processes operating on longer timescales via changing tropical Pacific background states. For example, CP El Ni\~nos are observed to occur more frequently after the $20^{th}$ century \cite{lee2010increasing}. Some work \cite{yeh2009nino} attributed this to the anthropogenic global warming. Others \cite{mcphaden2011nino}, however, suggested that the background-state changes observed in the tropical Pacific in the 2000s, i.e., a La Ni\~na-like pattern with enhanced trade winds and a more tilted thermocline, are opposite from those expected to produce more frequent CP El Ni\~no events. Based on this, it is argued that such a La Ni\~na-like, i.e., a strengthening Walker circulation, background state in the Pacific may favor the generation of CP El Ni\~no by suppressing convection and low-level convergence in the CP,  which could shift the anomalous convection westward \cite{xiang2013new}.

It should be noted that although the general circulation models (GCMs) are expected as the most ideal tool to investigate the ENSO complexity, it is still a great challenge for them to successfully simulate these ENSO characteristics. Besides, because GCMs include many factors that can influence ENSO, it is not always easy to uncover the physical processes behind model simulations. On the contrary, constructing a stochastic multiscale conceptual model that can depict the main features of all the interannual skeleton of ENSO, the intraseasonal wind bursts and the background Walker circulation simultaneously may be a promising way to understand the causes of ENSO complexity, which is the motivation of this work.

\begin{figure}[h]
\centering
\hspace{-1.5cm}\includegraphics[width=16cm,angle=0]{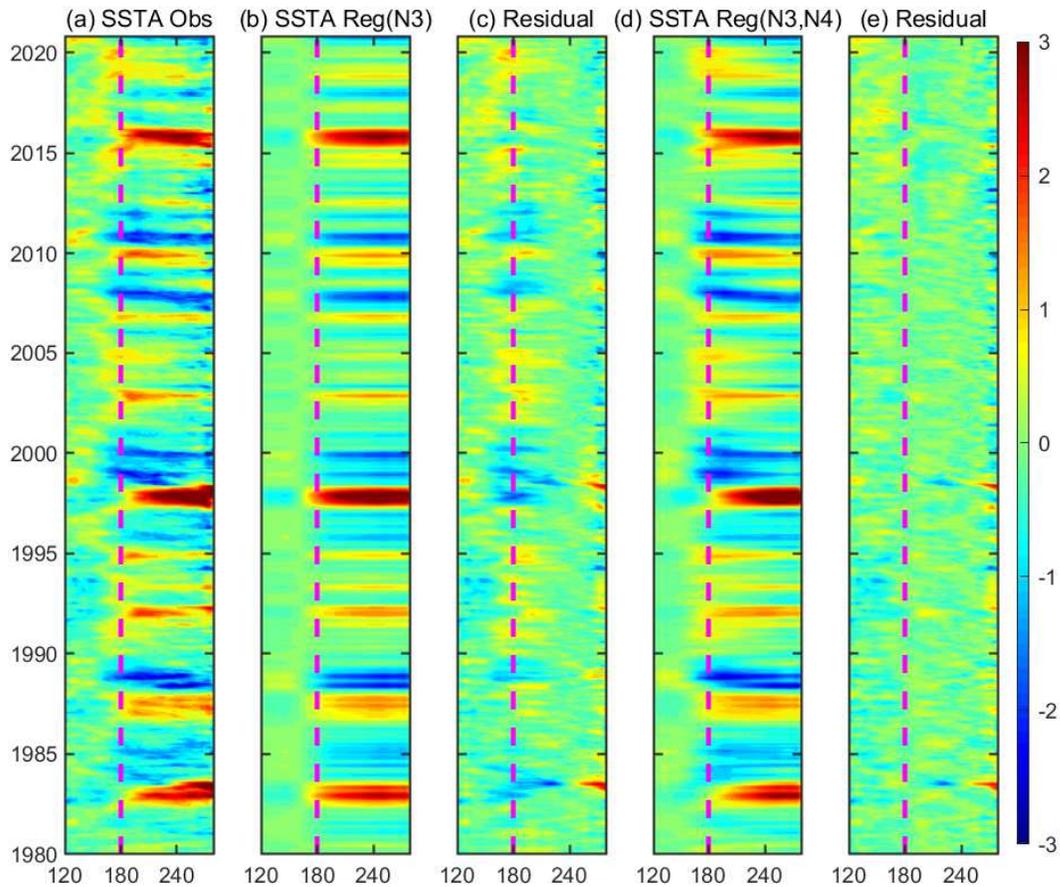}
\caption{Comparison between the univariate and bivariate linear regression models on reconstructing the observational SST variations in the equatorial Pacific. Panel (a): the original spatiotemporal evolutions of the SST anomaly field. Panel (b): the reconstructed SST anomaly field using the univariate linear regression based on $T_E$. Panel (c): the residual between the SST fields in Panel (a) and Panel (b). Panel (d): the reconstructed SST anomaly field using the bivariate linear regression based on $T_E$ and $T_C$. Panel (e):  the residual between the SST fields in Panel (a) and Panel (d).}
\label{plot_hov_reg_obs}
\end{figure}

On the subject of the interannual skeleton of ENSO, we need a model that can depict both the CP and EP SST anomalies, which are indispensable to simulate the ENSO complexity. To illustrate this point, Figure \ref{plot_hov_reg_obs}a shows the observational SST anomalies in the equatorial Pacific (averaged over $5^o$S-$5^o$N), along which are the regressed one with only the Ni\~no3 index (i.e., corresponding to EP in the model; Figure \ref{plot_hov_reg_obs}b) and that with both the Ni\~no3 and the Ni\~no4 (i.e., corresponding to CP in the model) indices (Figure \ref{plot_hov_reg_obs}d). These results clearly indicate that although the Ni\~no3-based univariate linear regression model captures many characteristics of the ENSO variation in the EP region, it fails to realistically depict those in the CP region. In fact, the CP events in 1991, 1995, 2003-2005 and 2020 are completely missed (see Figure \ref{plot_hov_reg_obs}b and the error plot in Figure \ref{plot_hov_reg_obs}c). In addition, due to the use of a solo SST variable in the regression, the eastward and westward propagating features (characterized by the underlying equatorial Kelvin and Rossby waves) are lost in the reconstructed ENSO spatiotemporal field. As a result, the reconstructed ENSO field contains only the standing oscillations from the central to the eastern Pacific. In contrast, the bivariate linear regression model significantly overcomes these shortcomings (Figure \ref{plot_hov_reg_obs}d). It succeeds in reproducing almost the same SST variation as is observed in nature in both the CP and EP regions (see the error plot in Figure \ref{plot_hov_reg_obs}e). The bivariate linear regression also facilitates the recovery of the large-scale behavior of the wave propagations across the equatorial Pacific, which allows the reconstructed SST field to highly resemble the observations.

Note that the physical mechanisms of the CP and EP El Ni\~no are quite different \cite{kao2009contrasting,kug2009two}. Specifically, due to the fact that the anomalous warming center of EP type of ENSO is located in the eastern Pacific, the mean thermocline is shallow and permits the perturbations on the subsurface to effectively influence the SST through upwelling processes. On the other hand, for the CP type of ENSO, the major warming center is concentrated in the central Pacific, where the zonal mean SST gradient is strongest due to the warm pool to the west and the cold tongue to the east, the anomalous zonal-current-related zonal advective feedback thus plays a very important role. Physical models of different degrees of complexity have also confirmed the important role of zonal advective feedback in causing the ENSO complexity \cite{kug2010warm,ham2012well,fang2015cloud,fang2018simulating}. Based on the above evidences, it is clear that including two degrees of freedom of the SST variation in a model, accounting for the CP and EP SST anomalies respectively, is essential to depict the large-scale features of the ENSO complexity.

This article aims at developing a three-region multiscale stochastic conceptual model for the ENSO complexity to
capture the general properties of the observed ENSO events as well as the complexity in patterns (e.g., Central Pacific vs. Eastern Pacific events), intensity (e.g., 10-20 year reoccurrence of extreme El Niños), and temporal evolution (e.g., more multi-year La Niñas than multi-year El Niños). It also aims at 
reproducing the observed non-Gaussian statistics in various Ni\~no regions, e.g., the positively skewed fat-tailed probability density function (PDF) in the Ni\~no3 region and the negatively skewed thin-tailed PDF in the Ni\~no4 region, which facilitates the model to quantify the uncertainty and capture the extreme events in the ENSO dynamics.

\section*{Results}


\subsection*{The three-region multiscale stochastic model}
In this work, a deterministic three-region conceptual model with the zonal advective feedback \cite{fang2018three} is adopted as a starting model. It is a general extension of the classical recharge oscillator model \cite{jin1997equatorial} and depict the air-sea interactions over the entire western, central and eastern Pacific. Its main advantage is to efficiently describe the different SST variations in the CP and EP regions, which has been shown to be indispensable to simulate the ENSO complexity (Figure \ref{plot_hov_reg_obs}). Then, a simple stochastic process describing the tropical atmospheric intraseasonal wind disturbances of the WWBs, the EWBs and the MJO, which involves a multiplicative noise that describes the modulation of the wind bursts by the interannual SST, is incorporated into the starting model. Such a stochastic parameterization of the intraseasonal variability plays a crucial role in explaining the irregularity of the ENSO events. Furthermore, a simple but effective stochastic process describing the multidecadal variation of the background Walker circulation \cite{yang2021enso} is incorporated into the system to modulate the strength and the occurrence frequency of the EP and CP El Ni\~nos.

\begin{figure}[h]
\centering
\hspace{-1.5cm}\includegraphics[width=16cm,angle=0]{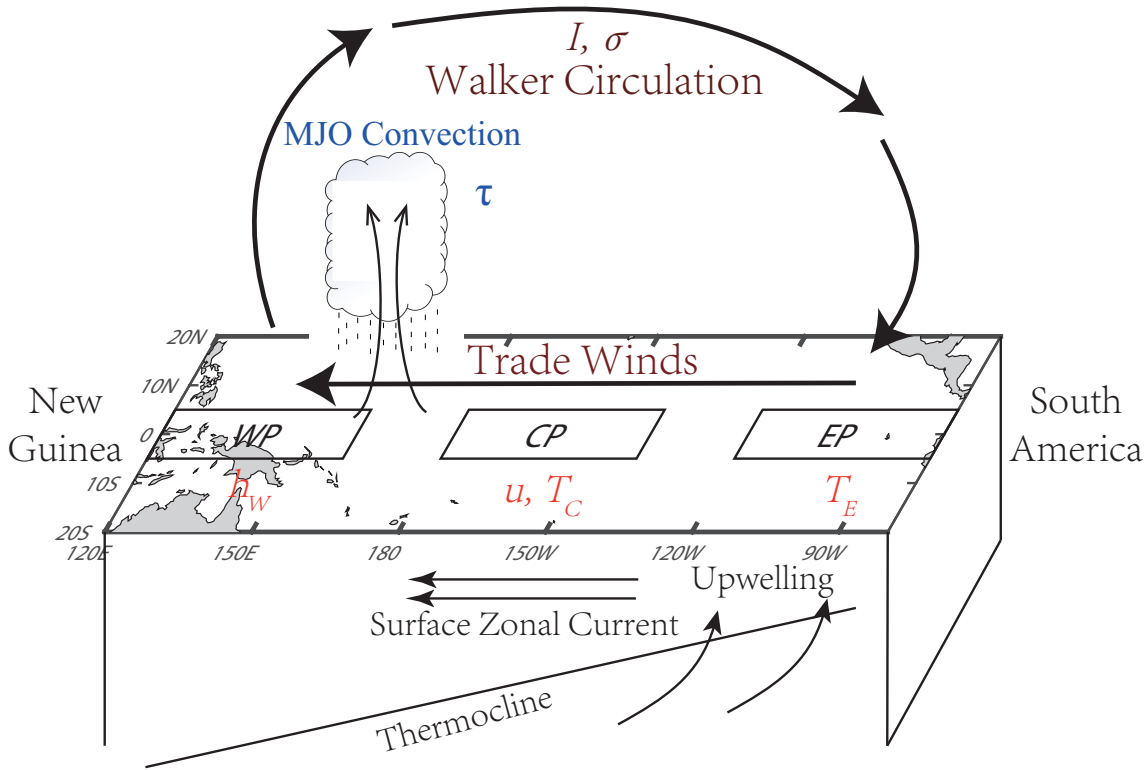}
\caption{Schematic diagram describing the main components of the three-region multiscale stochastic model. Specifically, it includes the interannual model that depicts the air-sea interactions over the entire western ($h_W$), central ($u, T_C$) and eastern Pacific ($T_E$), which is indispensable to simulate the ENSO complexity, the intraseasonal model that represents the random wind bursts and MJO ($\tau$), and the decadal model that illustrates the variation in the background strength of the Pacific Walker circulation ($I$) and the related zonal advective feedback ($\sigma$).}
\label{schematic}
\end{figure}

The three-region multiscale stochastic model is summarized as follows, where the details are described in the Methods section. The main components of this model and the multiscale interactions are also summarized in a schematic diagram as shown in Figure \ref{schematic}. The model reads,

	\begin{subequations}\label{Model_Stochastic}
		\begin{align}
			\frac{d u}{d t} &= -r u - \frac{\alpha_1 b_0 \mu}{2} (T_C + T_E) + {\beta_u\tau} + \sigma_u\dot{W}_u,\label{Model_Stochastic_u}\\
			\frac{d h_W}{d t} &= -r h_W - \frac{\alpha_2 b_0 \mu}{2} (T_C + T_E) + {\beta_h\tau} + \sigma_h\dot{W}_h,\label{Model_Stochastic_h_W}\\
			\frac{d T_C}{d t} &= \left(\frac{\gamma b_0\mu}{2} - c_1(T_C)\right) T_C + \frac{\gamma b_0 \mu}{2} T_E + \gamma h_W + \sigma u + C_u + \beta_C\tau +  \sigma_C\dot{W}_C,\label{Model_Stochastic_T_C}\\
			\frac{d T_E}{d t} &= \gamma h_W + \left(\frac{3\gamma b_0 \mu}{2} - c_2\right) T_E - \frac{\gamma b_0 \mu}{2} T_C + {\beta_E\tau}  + \sigma_E\dot{W}_E\label{Model_Stochastic_T_E}\\
			{\frac{d \tau}{d t}} & {= -d_\tau \tau + \sigma_\tau(T_C) \dot{W}_\tau},\label{Model_Stochastic_tau}\\
			{\frac{d I}{d t}} & {= -\lambda (I - m) + \sigma_I(I) \dot{W}_I}.\label{Model_Stochastic_I}
		\end{align}
	\end{subequations}
where the interannual component (Equations \eqref{Model_Stochastic_u}-\eqref{Model_Stochastic_T_E}) depicts the deterministic dynamics for both the CP and EP types of ENSO, the intraseasonal component represents the random wind bursts (Equation \eqref{Model_Stochastic_tau}), and the decadal component represents the variation in the background strength of the Pacific Walker circulation (Equation \eqref{Model_Stochastic_I}). In \eqref{Model_Stochastic}. $T_C$ and $T_E$ are the SST in the CP and EP while $u$ is the ocean zonal current in the CP and $h_W$ is the thermocline depth in the western Pacific (WP). The other two variables $\tau$ and $I$ represent the intraseasonal random wind burst amplitude, including the MJO, and the background Walker circulation, respectively. The decadal variability $I$ also stands for the zonal SST difference between the WP and CP regions that directly determines the strength of the zonal advective feedback. Besides, \eqref{Model_Stochastic} is an anomaly model, which means that all the prognostic variables are the deviations from their corresponding monthly climatology during the analysis period (years 1980-2020).	

Stochasticity plays a crucial role in coupling variables at different time scales and parameterizing the unresolved features in \eqref{Model_Stochastic}. First, the intraseasonal component $\tau$ is modeled by a simple stochastic differential equation \eqref{Model_Stochastic_tau} with a state-dependent (i.e., multiplicative) noise coefficient $\sigma_\tau$, where $\dot{W}_\tau$ is a white noise source. The stochastic wind bursts are then coupled to the processes of the interannual variables serving as external forcings, which are the main mechanism for generating the EP events and the non-Gaussian features of $T_E$. In addition to the stochastic wind bursts, four Gaussian random noises  $\sigma_u\dot{W}_u$, $\sigma_h\dot{W}_h$, $\sigma_C\dot{W}_C$ and $\sigma_E\dot{W}_E$ are further added to the processes describing the interannual variabilities. These random forcings effectively parameterize the additional contributions to the interannual variables that are not explicitly modeled here, such as the subtropical atmospheric forcing \cite{fang2020control}. Next, the background Walker circulation in the decadal time scale has been shown to modulate the interannual variability \cite{mcgregor2014recent,chen2015strong}. Since the details of the background Walker circulation consist of uncertainties and randomness, a simple but effective stochastic process is used to describe the time evolution of the decadal variability $I$, where $\dot{W}_I$ in \eqref{Model_Stochastic_I} is a white noise source. The multiplicative noise in the process of $I$ aims at guaranteeing the positivity of $I$ due to the fact that the long-term average of the background Walker circulation is non-negative.

The coupled model \eqref{Model_Stochastic} involves a minimum nonlinearity, which nevertheless plays a crucial role in recovering the key dynamics and reproducing the non-Gaussian statistics for the CP events. The first nonlinearity is the $\sigma u$ term in \eqref{Model_Stochastic_T_C}, which represents that the strength of the zonal advection is modulated by the decadal variability. Such a nonlinearity is crucial in simulating the right occurrence frequency of both the CP and EP El Ni\~no events. Another key nonlinearity comes from the coefficient $c_1$ in \eqref{Model_Stochastic_T_C}, which is a quadratic function of $T_C$. In other words, the total damping in \eqref{Model_Stochastic_T_C} is cubic. Such a cubic nonlinearity is justified by analyzing the observational data (see Figure \ref{c1_estimate} and the detailed justifications in the Methods section). It also facilitates the recovery of the non-Gaussian statistics in the CP region, which has completely different characteristics as that in the EP region. Note that, since the coupled model is nonlinear, the long-term statistics does not necessarily have a zero mean. To guarantee the model \eqref{Model_Stochastic} to be an anomaly model, an extra $C_u$ term is imposed in \eqref{Model_Stochastic_T_C} such that all the variables have climatology with zero mean.

Finally, seasonal phase locking is a remarkable feature of the ENSO, which manifests in the tendency of ENSO events to peak during boreal winter \cite{tziperman1997mechanisms, stein2014enso}. Here, the effects of seasonality are added to both the wind activity and the collective damping. The former accounts for the active phase of the MJO in boreal winter \cite{zhang2005madden}  while the latter is due to the seasonal migration of the Intertropical Convergence Zone (ITCZ), which modulates the strength of the upwelling and horizontal advection processes to influence the evolution of the SST \cite{mitchell1992annual}. Thus, the three coefficients $c_1, c_2$ and $\sigma_\tau$ are all time-periodic functions.

The dimensional units and the parameters in the coupled model \eqref{Model_Stochastic} are summarized in Table \ref{Table_Parameters}. As is described in Methods, all parameters are determined  based on the observational data during 1980-2020.

\subsection*{Numerical results}
\subsubsection*{Reproducing the observed ENSO statistics}
\begin{figure}[h]
\centering
\hspace*{-2cm}\includegraphics[width=18cm,angle=0]{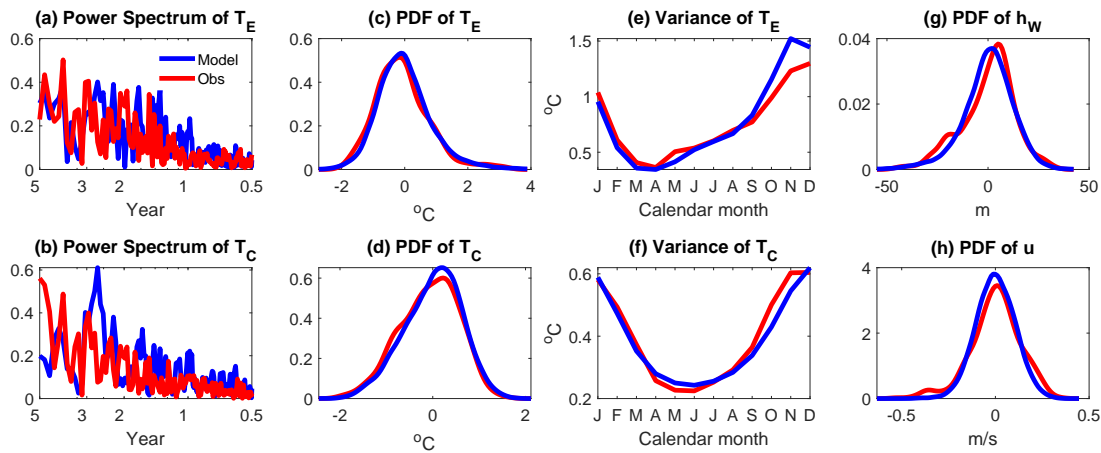}
\caption{Comparison of the statistics of the observations and the coupled multiscale stochastic model \eqref{Model_Stochastic}. Panels (a)--(b): power spectrums of Ni\~no3 and Ni\~no4 SST. Panels (c)--(d): PDFs of Ni\~no3 and Ni\~no4 SST. Panels (e)--(f):
the monthly variance (i.e., the seasonal cycle) of Ni\~no3 and Ni\~no4 SST. Panel (g): PDF of the thermocline depth $h_W$ in the  western Pacific region. Panel (h): PDF of the ocean zonal current $u$ in the CP region. In each panel, red and blue curves are for the observation and model, respectively. All the statistics of the model are computed based on a $2000$-year long simulation.}
\label{PDFs_Spectrums}
\end{figure}

Figure \ref{PDFs_Spectrums} compares several key statistics of the multiscale stochastic model \eqref{Model_Stochastic} with those from the observations. The model statistics is based on a 2000-year simulation, which is long enough to provide unbiased results. Simulations with different random number seeds have been utilized to confirm the robustness of the statistics.

Figures \ref{PDFs_Spectrums}a–b  show the power spectrums of the Ni\~no3 and the Ni\~no4 SST, respectively. Except a slight overestimation of the SST spectrum in the CP region around the frequency of 3 years, the model recovers the spectrums of both the Ni\~no3 and Ni\~no4 SST in a quite accurate fashion. Such a result indicates that the model is able to reproduce the observed irregular oscillations in both the Ni\~no regions.

Next, Figures \ref{PDFs_Spectrums}c–d illustrate that the model perfectly recovers the remarkable non-Gaussian statistics of both the Ni\~no3 and Ni\~no4 SST. In particular, the observed Ni\~no3 SST has a positive skewness and a one-sided fat tail that results from the occurrence of the extreme El Ni\~no events. Due to the multiplicative noise in the wind burst process \eqref{Model_Stochastic_tau}, the model is able to accurately reproduce such a highly non-Gaussian statistical feature. On the other hand, the skewness of the observed Ni\~no4 SST is negative, and the kurtosis is 2.7, which is less than the standard Gaussian value 3, indicating the suppression of extreme El Ni\~no events in the CP region. Thanks to the cubic and non-centered damping $c_1$ (Equation \eqref{c1_equation} in the Methods section), the model succeeds in capturing such a skewed and light tailed distribution. Note that GCMs and the intermediate models often have great difficulties in reproducing these highly non-Gaussian PDFs, which are nevertheless one of the most important and necessary conditions for simulating the realistic ENSO complexity.

In addition to reproducing the climatology distribution functions, the model is also skillful in recovering the observed seasonal phase locking features of the EP and the CP ENSO. This can be seen in Figures \ref{PDFs_Spectrums}e–f , which show the monthly variance of the Ni\~no3 and the Ni\~no4 SST, respectively. Both types of ENSO onset in boreal spring, develop in summer, and peak in the following winter. The model also realistically reproduces the slight late onset (about two months) of the CP ENSO than the EP ENSO. The late onset reduces the growing season \cite{xu2001role} and is a key reason why the CP events are typically weaker (i.e., smaller variance) than the EP events \cite{yu2017changing}.

Finally, Figures \ref{PDFs_Spectrums}g–h show that the model is able to recover the observed variance of $h_W$ and $u$ as well. This indicates the skill of the model in quantifying the uncertainty of nature.

\subsubsection*{Reproducing the observed ENSO complexity}
ENSO complexity appears in its spatial pattern, peak intensity, and temporal evolution. Table \ref{Table_complexity} compares the model results for different situations with the observation on the ENSO complexity. It also summarizes the results of the sensitivity experiments in the next subsection.

In term of the complexity in the ENSO pattern, this multiscale stochastic model produces 660 El Ni\~nos and 852 La Ni\~nas during its 2000-year simulation based on a widely used ENSO classification method \cite{yu2018distinct}. Among the El Ni\~no events, about 60\% (398 events) of them are EP events and 40\% (262 events) of them are CP events. Note that during the observational period from 1950 to 2020, 14 of the 24 (i.e., 58\%) major El Ni\~nos are of the EP type, and 10 of them (i.e., 42\%) are of the CP type \cite{yu2012changing}. Such a comparison indicates that the model reproduces roughly the same ratio of the EP and CP events as the observations. 

In term of the complexity in the ENSO intensity, there is a tendency for extreme El Ni\~no events (e.g., the 1997-98 and 2015-16 ones) to occur every 10-20 years as in the observations \cite{sun200910}.   A total of 4 extreme El Ni\~no events have occurred since 1950, namely 1972-73, 1982-83, 1997-98 and 2015-16. Consistent with this reoccurrence frequency, the multiscale stochastic model produces 125 extreme El Ni\~no events in its 2000 simulation (namely, on average every 16 years). 

In term of the complexity in the ENSO evolution, it is noted that an El Ni\~no (La Ni\~na) event can be followed by a La Ni\~na (El Ni\~no) event to result in a cyclic ENSO evolution pattern or by another El Ni\~no (La Ni\~na) event to result in a multi-year ENSO evolution pattern \cite{yu2018distinct}. Multi-year ENSO events are a major challenge for the accurate ENSO prediction \cite{xue2013prediction}. During the historic period, multi-year La Ni\~na events tend to occur more frequently than multi-year El Ni\~no events \cite{fang2020contrasting}.
However, the GCMs were often not able to reproduce this asymmetric feature \cite{fang2020contrasting}. Such a  deficiency in the operational models, together with the limited number of multi-year ENSO events in the observations, have hindered the effort to uncover the underlying dynamics of multi-year ENSO and its associate El Ni\~no-La Ni\~na asymmetry. In contrast, it is very encouraging to find that the multiscale stochastic model developed here is able to produce more multi-year La Ni\~nas (209) events than multi-year El Ni\~nos (100) events in its 2000-year simulation. Thus, this model can be a useful tool to better understand the multi-year ENSO dynamics. 

\begin{figure}[h]
\centering
\hspace{-0.5cm}\includegraphics[width=16cm,angle=0]{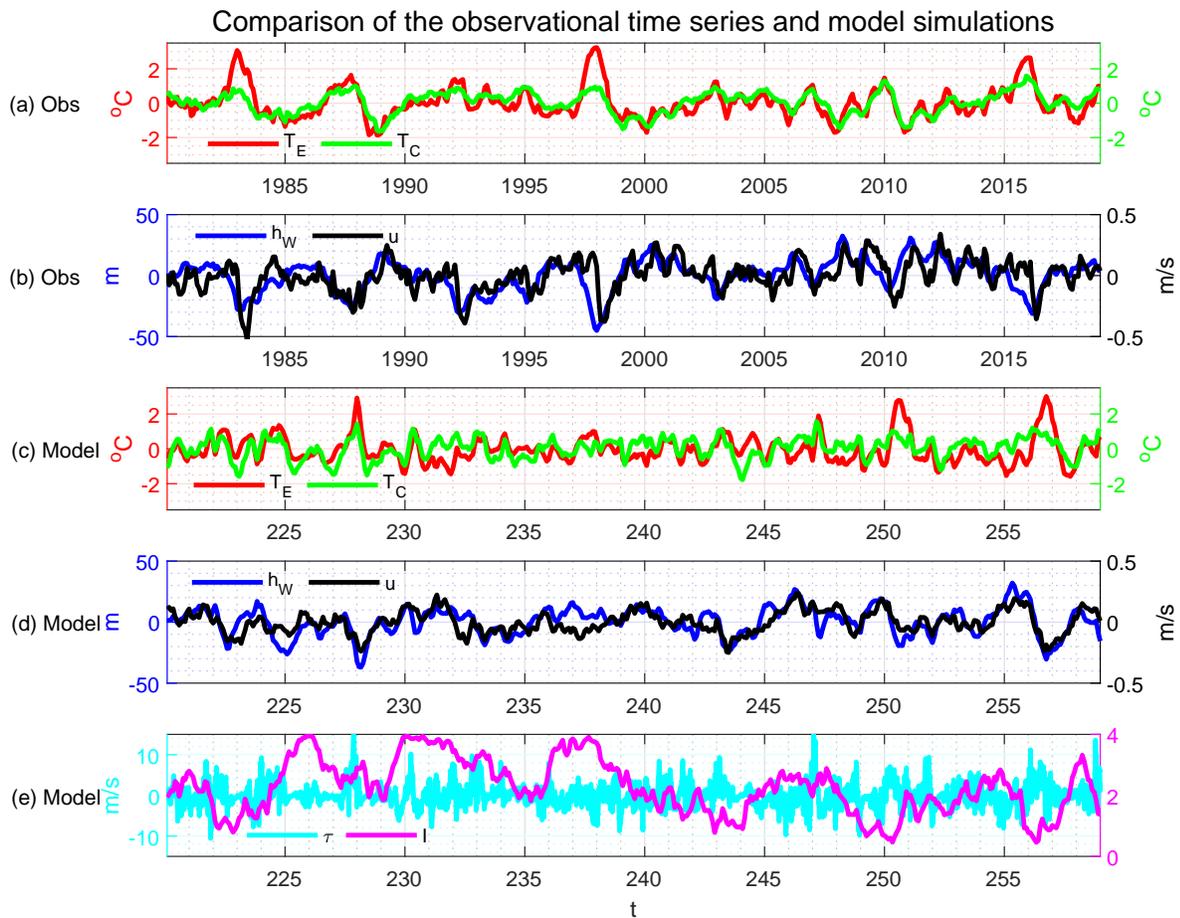}
\caption{Comparison of the observational time series and the model simulations. Panels (a)--(b): the observational SST anomalies in the Ni\~no3 (red) and Ni\~no4 (green) regions, the observed thermocline depth anomaly in the western Pacific region (blue) and the observed ocean zonal current in the CP region (black). Panels (c)--(d): similar to Panels (a)--(b) but for the results from the model. Panel (e): the time series of the intraseasonal wind bursts $\tau$ (cyan) and the decadal index $I$ (purple) from the model.}
\label{Model_Obs_TimeSeries}
\end{figure}

To use the model simulation for a better understanding of the physical processes behind the ENSO complexity, Figure \ref{Model_Obs_TimeSeries} compares key atmospheric and ocean variables during a particular 40-year segment of the model simulation with those observed during the past four decades (1980-2020). Both the observed and simulated $T_E$ and $T_C$ indices (Figures \ref{Model_Obs_TimeSeries}a and \ref{Model_Obs_TimeSeries}c) clearly indicate that most of the extreme El Ni\~no events are of the EP type. For these extreme events, the amplitude of the Ni\~no4 SST (i.e., $T_C$) is significantly smaller than that of its counterpart Ni\~no3 SST (i. e., $T_E$). This is consistent with the finding from the observations that the vertical thermocline process produces strong El Ni\~no events (i.e., the EP El Ni\~nos), while the zonal advection process produces weak El Ni\~no events (i.e., the CP El Ni\~no). In the simulation, the time series of $T_E$ and $T_C$ are positively correlated with each other and the positive correlation is also found between $u$ and $h_W$. The latter two, on the other hand, have negative correlations with the formers, which provide the delayed negative feedbacks according to the recharge oscillator theory. It is noticed that extreme El Ni\~no events are preferable when the decadal variable $I$ is close to zero (Figure \ref{Model_Obs_TimeSeries}e). Specifically, under such circumstances, the model has a high chance to generate strong WWBs and therefore more moderate and extreme EP events are likely to be triggered. In contrast, when $I$ becomes large, the warming center tends to occur in the CP region. Consistent with the observations \cite{yu2012changing}, the slow variation of the background strength of the Pacific Walker circulation modulates the occurrence frequencies of the EP and CP ENSOs. 

\begin{figure}[h]
\centering
\hspace{-0.5cm}\includegraphics[width=16cm,angle=0]{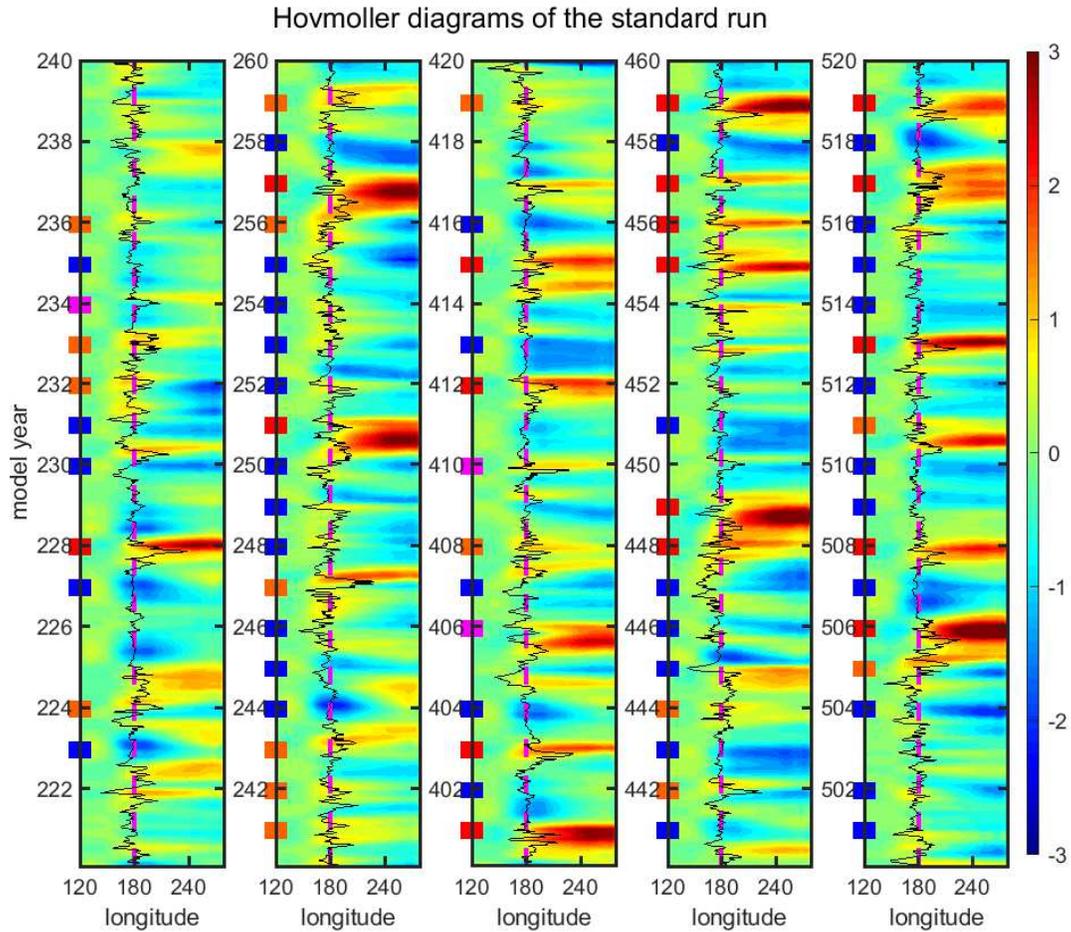}
\caption{Hovmoller diagrams based on the model simulations from the standard run, i.e., using the parameters in Table \ref{Table_Parameters}. The equatorial SST variations are obtained by the bivariate linear regression. Here the coefficients of the regression model are obtained using the observational data. Then the Ni\~no3 and Ni\~no4 indices $T_E$ and $T_C$ from the model are plugging into the regression model to obtain the SST spatiotemporal patterns.
The colored boxes on the left vertical axis (ranging from September to the next February) indicate the types of the ENSO events in boreal winter, which are based on the definitions in Methods section. The red, purple, orange and blue boxes are for the strong EP El Ni\~no, the moderate EP El Ni\~no, the CP El Ni\~no and the La Ni\~na events, respectively. The wind burst time series is placed on top of the Hovmoller diagram. The center of the wind burst time series is located at the dateline, where the WWBs and EWBs correspond to the time series values going towards the right and left from the dateline. The distance from the dateline represents the strength and direction of the wind bursts.}
\label{Model_hov_particular_events}
\end{figure}

Another advantage of the model developed here is that it can be combined with the bivariate regression method to reconstruct the spatiotemporal evaluation of the SST field,

\begin{equation}\label{regression}
\mbox{SST}(x,t) = r_C(x)T_C(t) + r_E(x)T_E(t)
\end{equation}
which provides a clearer view of understanding the ENSO complexity dynamics from the model. Here $x$ is the longitude and $t$ is time. The regression coefficients $r_C(x)$ and $r_E(x)$ are determined using the observational data at each longitude grid point $x$. Then the Ni\~no3 and Ni\~no4 indices $T_E$ and $T_C$
from the model are plugging into the regression formula \eqref{regression} to obtain the SST spatiotemporal patterns.
Figure \ref{Model_hov_particular_events} shows the Hovmoller diagrams of the model simulation, including the $40$-year period in Figure \ref{Model_Obs_TimeSeries}, which clearly demonstrate the ENSO complexity.\\
\textbf{Extreme El Ni\~no.} First, the model is able to simulate realistic extreme El Ni\~nos (red color), mimicking the observations \cite{sun200710}. Some examples of the extreme El Ni\~no events are those in years 228, 251, 401 and 459, where the associated SST patterns and the profile of their precursors, i.e., the wind burst amplitudes and directions, are all similar to the observed 1987-1988 and 1997-1998 events.\\
\textbf{Delayed Extreme El Ni\~no.} Notably, the model is able to simulate the so-called delayed extreme El Ni\~no as was observed in 2015-2016, for example, the two events during years 448-449 and 505-506 in Figure  \ref{Model_hov_particular_events}. The reason for the model to generate this kind of El Ni\~no \cite{thual2019statistical} is its success in simulating the associated peculiar WWB-EWB-WWB structure \cite{hu2016exceptionally}. Here, the first WWB tends to trigger a strong El Ni\~no but the following EWB kills the event and postpones it until the next year when another series of the WWBs occur.\\
\textbf{Moderate EP El Ni\~no.} Next, the model is able to simulate the traditional moderate EP El Ni\~no (purple color; e.g., years 225, 238, 403 and 445), which are triggered by the moderate WWBs.\\
\textbf{Isolated and Consecutive CP El Ni\~no.} In addition to the complexity of the EP events, one desirable feature of the model is that the simulated CP El Ni\~no events (orange color) also highly resemble the observations. In particular, both a single CP event (e.g., years 222, 259 and 444) and a sequence of consecutive CP events (e.g., years 232-233 and 241-243) can be reproduced from the model. The latter mimics the observed CP episodes during 2003-2006. \\
\textbf{Mixed CP-EP events.} In addition, the model can create some mixed CP-EP events (e.g., years 241, 247 and 457), which are similar to the observed ones in early 1990’s.\\
\textbf{Single- and multi-year La Ni\~na.} Finally, the La Ni\~na events from the model (blue color; e.g., years 223, 404 and 504) usually follow the El Ni\~no ones. Some La Ni\~na events have cold SST in the CP region while other La Ni\~na years have cold centers locating around the EP area. The model is also able to generate multi-year La Ni\~na events, i.e., a La Ni\~na transitions to another La Ni\~na, such as the one spans over year 509 and year 510.

\subsubsection*{Sensitivity analysis}
The simple formula of this multiscale stochastic model and their key parameters also enable us to project the possible changes of ENSO complexity under various past and future climate regimes. Several sensitivity tests are utilized for a further understanding of the coupled multiscale stochastic model \eqref{Model_Stochastic}.

The first study is on the damping coefficient $c_1$ in \eqref{Model_Stochastic_T_C}, which reflects the collective residual part of the heat budget equation apart from the dynamical terms.  Recall that a cubic damping is adopted in Table \ref{Table_Parameters}, i.e., $c_1$ is small for the small local SST and becomes larger for the large SST anomalies. Such a treatment avoids the unrealistic enlargement of the simulated SST anomalies \cite{jin1997equatorial}. If a linear damping is utilized, then the major change is the kurtosis of $T_C$, which will become larger than 3 and lead to fat tails of the Ni\~no4 PDF. Despite that the overall model simulation remains similar, there are occasionally certain CP events that have large amplitudes, which is not the case in the observations. This indicates the necessity of adopting a cubic damping in $c_1$ to capture the observed non-Gaussian features in the CP region.

Next, if an additive noise $\sigma_\tau$ is used in the wind burst equation $\tau$ \eqref{Model_Stochastic_tau}, i.e., $\sigma_\tau$ being independent with the variations on the interannual timescale, then the PDF of $T_E$ will become nearly Gaussian. As a consequence, the occurrence of the extreme El Ni\~no events will become much less frequent and the amplitudes of the La Ni\~na will become stronger. This reflects the importance of the observational character, i.e., there being a deterministic part of the wind bursts modulated by the low‐frequency SST variation associated with ENSO, in triggering the asymmetry of the EP El Ni\~no.

\begin{figure}[h]
\centering
\hspace{-0.5cm}\includegraphics[width=16.5cm,angle=0]{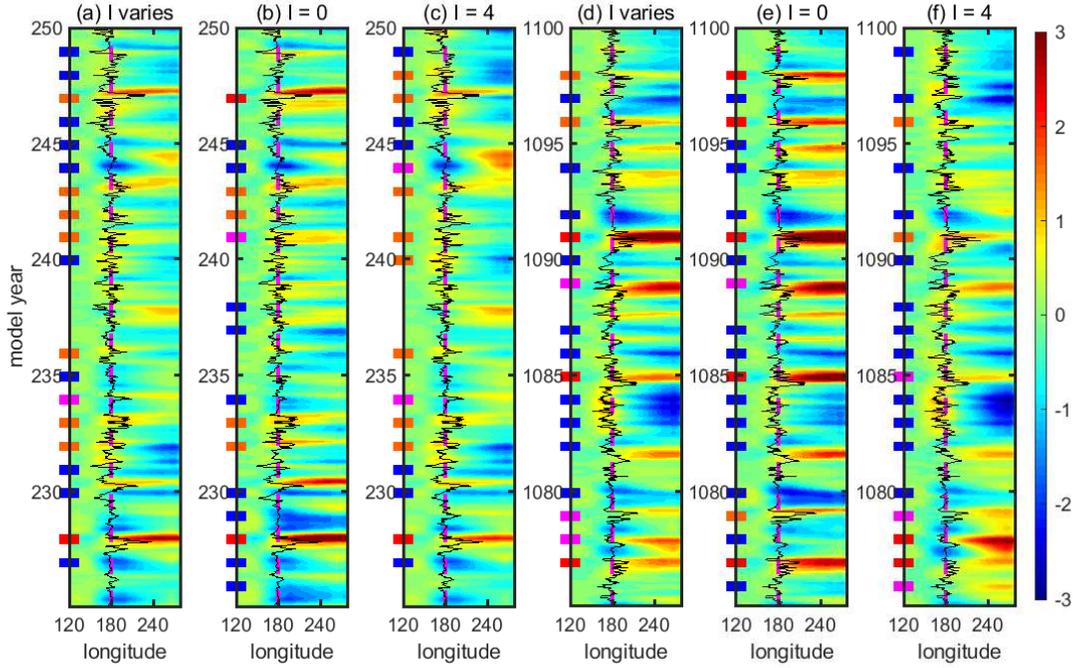}
\caption{Comparison of the standard run of the model (Panels (a) and (d)), the simulation with  $I\equiv 0$ (Panels (b) and (e)) and the simulation with a constant $I\equiv4$ (Panels (c) and (f)). For a fair comparison, the random number generators $\dot{W}_\tau$ in the wind burst equation \eqref{Model_Stochastic_tau} in the three cases are set to be the same. }
\label{Model_hov_I0_I4}
\end{figure}

Finally, the role of the decadal variability $I$ is studied in Figure \ref{Model_hov_I0_I4}. Here, in addition to the standard run with a time-varying $I$ as in \eqref{Model_Stochastic_I}, the other two tests both have a fixed value of $I$, with $I\equiv0$ and $I\equiv4$, respectively. For a fair comparison, the random number generators $\dot{W}_\tau$ in the wind burst equation \eqref{Model_Stochastic_tau} in the three cases are set to be the same.

Clearly, if $I\equiv0$, i.e., the background Walker circulation and zonal thermocline slope is relatively weak, then CP events occur less frequently while EP events become dominant (Table \ref{Table_complexity}).  Note that, even with $I\equiv0$ in such a case, there remains a small number of the CP El Ni\~nos in the simulation, which are triggered by the stochastic noise. Quantitatively, in this case, 646 El Ni\~nos and 861 La Ni\~nas can be identified in the 2000 model years. Among the El Ni\~nos, 453 events are EP type and 193 events are CP one. Also, less multi-year ENSO events are generated, i.e., 78 multi-year El Ni\~nos and 200 La Ni\~nas in total being found. Besides, there are 260 extreme El Ni\~no events in total, i.e., a tendency of its occurrence for every 8 years. This frequency is twice as the standard run, consistent with the observation that 3 out of the 4 extreme El Ni\~no events occurred before 2000. This is also consistent with the projection that an increasing frequency of extreme El Ni\~no events will emerge due to the greenhouse warming, since a projected surface warming over the EP that occurs faster than in the surrounding ocean waters \cite{cai2014increasing}.

On the other hand, if $I\equiv4$, i.e., with a relatively strong background Walker circulation and zonal thermocline slope, then the model simulations fail to capture many important features in observations (Table \ref{Table_complexity}). It is clear from Figure \ref{Model_hov_I0_I4} that many of the strong El Ni\~no in the Ni\~no3 region will disappear. Specifically, only 34 extreme El Ni\~no events can be identified in this situation, which is much less than that in the standard run (125). In addition, the occurrence of the CP El Ni\~no and La Ni\~na will be more frequent and the resulting PDF of $T_E$ will become nearly Gaussian, which is fundamentally different from the observed non-Gaussian PDF with a fat tail.



 
\section*{Discussion}\label{Sec:Conclusion}
As was shown in the context, a three-region multiscale stochastic model is developed to show that the observed ENSO complexity can be explained by combining intraseasonal, interannual, and decadal processes. The model starts with a deterministic, linear and stable system for the interannual variabilities, which includes both the ocean heat content discharge/recharge and the ocean zonal advection. Then two stochastic processes with multiplicative noise describing the intraseasonal wind bursts and the decadal variation of the Walker circulation are incorporated. This three-region multiscale stochastic model can reproduce not only the general 
properties of ENSO events observed during the period of 1980-2020, but also the observed complexity in ENSO patterns (e.g., CP vs. EP El Ni\~nos), intensity (e.g., $\sim$10-20 year reoccurrence frequency of extreme El Ni\~no events), and evolution patterns (e.g., more multi-year La Ni\~nas than multi-year El Ni\~nos), which are often hard to be simulated by the state-of-the-art models. The model also perfectly recovers the non-Gaussian SST statistics of nature in reproducing both the positively skewed fat-tailed PDF in the Ni\~no3 region and the negatively skewed thin-tailed PDF in the Ni\~no4 region, which allows a systematic uncertainty quantification of the ENSO dynamics and facilitates the study of the extreme El Ni\~no events.

Except for the stochasticity of the model, the nonlinearity also plays an important role. In fact, based on a heat budget analysis of the mixed layer temperature with the observational dataset, the collective damping rate over the CP region is parameterized as a cubic polynomial function in terms of $T_C$. This is found to be crucial for obtaining the realistic negative-skewed PDF for the simulated $T_C$ and therefore for simulating the realistic ENSO complexity. It should also be noted that the theoretical explanations of ENSO are always grouped into two categories \cite{timmermann2018nino,zebiak1987model,wang2018review}. In the first category, ENSO is viewed as a self-sustained, unstable and naturally oscillatory mode of the coupled ocean-atmosphere system, in which the nonlinearity acts mainly to bound the growing eigenmode and create a finite amplitude of the ENSO cycle. In the other category, ENSO is regarded as a stable (damped) mode triggered by atmospheric random "noise" forcing. Based on the results of this work, we conclude that the first explanation is more suitable for depicting the CP type of ENSO, since the nonlinearity plays an important role for its evolution. On the other hand, the second theory is more appropriate for explaining the development of the EP type of ENSO. In fact, the stochastic forcings, i.e., the WWBs (EWBs), are crucial for both the occurrences and the amplitudes of the EP El Ni\~no (La Ni\~na). This indicates that both the nonlinearity and stochastic processes are of great importance for simulating and studying the ENSO complexity.

It has been shown in this paper that the $T_C$ and $T_E$ time series combined with the bivariate regression technique can be utilized to simulate the spatiotemporal patterns of the SST, which are qualitatively and quantitatively similar to the observations. The stochastic forcing in the conceptual model allows the reconstructed spatiotemporal patterns to have the same level of the irregularity as nature, which outweighs most of the GCMs that are more deterministic. Thus, one direct application of this model is for prediction. On the one hand, the model itself is ready for efficient data assimilation and ensemble forecasts. On the other hand, the computational efficiency and the physical consistency of the model facilitates the machine learning forecast of ENSO. More specifically, the model can be easily used to create ENSO spatiotemporal patterns for several thousand years. Then the transfer learning technique \cite{torrey2010transfer} can be used to further improve the quality of these time series with the help of the limited but valuable observational data, which provides effective training data for the machine learning forecasts. Besides, the sensitivity analysis of the decadal variability I suggests that this model can be utilized to effectively and quantitatively analyze the influences of the background state changes, e.g., the greenhouse warming or the historical changes at millennial timescales, on the ENSO characters. Finally, the current modeling framework allows to further incorporate detailed additional physical processes for the ENSO complexity, such as the subtropical atmospheric forcing \cite{fang2020control}, which are now described by stochastic parameterizations.



\section*{Methods}\label{Sec:methods}

\subsection*{The datasets}
The monthly ocean temperature and current data used here are all from the GODAS dataset \cite{behringer2004evaluation}. The thermocline depth along the equatorial Pacific is approximated from the potential temperature as the depth of the $20^o$C isotherm. GODAS dataset is available at a horizontal resolution of $1/3^o\times1/3^o$ near the tropics and has 40 vertical levels with 10m resolution near the surface. The analysis period is from 1980 to 2020. Anomalies presented in this study are calculated by removing the monthly mean climatology of the whole period. In this work, the Ni\~no4 ($T_C$) and the Ni\~no3 ($T_E$) indices are the average of SST anomalies over the regions $160^o$E-$150^o$W, $5^o$S-$5^o$N and $150^o$W-$90^o$W, $5^o$S-$5^o$N, respectively. The $h_W$ index is the average of thermocline depth anomaly over $120^o$E-$180^o$, $5^o$S-$5^o$N while the $u$ index is the average mixed-layer zonal current in the CP region.

Next, the daily zonal wind data at 850 hPa from the NCEP–NCAR reanalysis \cite{kalnay1996ncep} is used to describe the intraseasonal wind bursts. After removing the daily mean climatology, the anomalies are projected to the WP region to create the wind burst index, which is shown in Figure \ref{Variable_I}a.

In addition to the interannual and intraseasonal data, the Walker circulation strength index is adopted to illustrate the modulation of the decadal variation on the interannual ENSO characters. It is defined as the sea level pressure difference over the central/eastern Pacific ($160^o$W-$80^o$W, $5^o$S-$5^o$N) and over the Indian Ocean/west Pacific ($80^o$E-$160^o$E, $5^o$S-$5^o$N)\cite{kang2020walker}. It should be stressed that the zonal SST gradient between the WP and CP region is highly correlated with this Walker circulation strength index (i.e., their simultaneous correlation coefficient is around $0.85$), suggesting the significant air-sea interacting characteristic over the equatorial Pacific. Since the latter is more directly related to the zonal advective feedback strength over the CP region, the decadal model mainly reflects this variation.

\subsection*{Definitions of different types of the ENSO events}
To quantify the ENSO complexity, the definitions of different El Ni\~no and La Ni\~na events are as follows, which are based on the average of the SST anomalies over the boreal winter (December-January-February). Following the definitions in the reference \cite{kug2009two}, when the EP is warmer than the CP and is warmer than $0.5^o$C, it is classified as the EP El Ni\~no. 
Following the definitions used by the reference \cite{wang2019historical}, an extreme El Ni\~no event corresponds to the situation that the maximum of EP SST anomaly from April to the next March is larger than $2.5^o$C. When the CP is warmer than the EP and the CP SST anomaly is larger than $0.5^o$C, the event is then defined as a CP El Ni\~no. Finally, when either the CP and EP SST anomaly is less than $-0.5^o$C, it is defined as a La Ni\~na event.

\subsection*{The starting interannual model: a deterministic three-region conceptual system} To develop the three-region multiscale stochastic model \eqref{Model_Stochastic},
the associated starting interannual model with four unknowns over the western, central, and eastern Pacific regions is as follows,

\begin{subequations}\label{Model_Deterministic}
\begin{align}
\frac{d u}{dt} &= -r u - \frac{\alpha_1 b_0 \mu}{2} (T_C + T_E),\label{Model_Deterministic_u}\\
\frac{d h_W}{dt} &= -r h_W - \frac{\alpha_2 b_0 \mu}{2} (T_C + T_E),\label{Model_Deterministic_h_W}\\
\frac{d T_C}{dt} &= \left(\frac{\gamma b_0\mu}{2} - c\right) T_C + \frac{\gamma b_0 \mu}{2} T_E + \gamma h_W + \sigma u,\label{Model_Deterministic_T_C}\\
\frac{d T_E}{dt} &= \gamma h_W + \left(\frac{3\gamma b_0 \mu}{2} - c\right) T_E - \frac{\gamma b_0 \mu}{2} T_C,\label{Model_Deterministic_T_E}
\end{align}
\end{subequations}
where $T_C$ and $T_E$ are the SST in the CP and EP, respectively, while $u$ is the ocean zonal current in the CP and $h_W$ is the thermocline depth in the WP. All the four variables are anomalies. This starting interannual skeleton model for ENSO is constructed to depict the air–sea interactions over the entire western, central and eastern Pacific \cite{fang2018three}. The key physics of the model are summarized as follows. First, the monthly variations of the thermocline slope and zonal wind stress over the central-western and the central-eastern Pacific regions are tightly linked through the Sverdrup balance relationships, as observed. As a result, if one obtains the variation of the thermocline depth anomalies over the WP ($h_W$), those over the CP ($h_C$) and EP ($h_E$) can also be diagnosed. Second, in the absence of 
the ocean zonal advection, the dynamic equations of the $h_W$ and $T_E$ degenerate to those in the recharge paradigm. Third, to introduce the zonal advective feedback, a simple equation for the mixed-layer zonal current is adopted. Finally, in contrast to the recharge paradigm, which considers the thermocline feedback as the only positive feedback in the EP, the development of the SST in the CP is also influenced by the zonal advective feedback. Combining these elements yields the linear coupled system \eqref{Model_Deterministic}.

It can be seen that when the coefficient $\sigma$ in \eqref{Model_Deterministic_T_C} is set to be zero, i.e., ignoring the zonal advective feedback, the system will degrade into the recharge paradigm, with $T_C=T_E$, which illustrates the EP type of ENSO with no emphasis on the differences between the CP and EP regions \cite{jin1997equatorial}. Therefore, the three-region model \eqref{Model_Deterministic} can be seen as an extension of the recharge paradigm. In \eqref{Model_Deterministic}, the collective damping rate $c$ is dominated by the time scale over which water in the equatorial band is replaced by the mean climatological upwelling, i.e., about $2$ months; the parameter $\gamma$ measures the strength of the thermocline feedback, which is chosen to give an SST rate of change of $1.5^o$C over $2$ months per $10$ m of the thermocline depth anomaly over the eastern Pacific. Similarly, the coefficient $\sigma$, which measures the strength of the zonal advective feedback, is chosen to give an SST rate of change of $1.5^o$C over $2$ months per $0.5$ m/s of the zonal current anomaly over the central Pacific, i.e., the background zonal SST difference between the WP and CP is $3^o$C. The collective damping rate $r$ in the ocean adjustment is set as $1/$($8$ months), which is induced by the loss of energy to the boundary currents of the west and east sides of the ocean basin. Due to the fact that, for a given steady zonal wind stress forcing, the zonal mean thermocline depth anomaly of the recharge oscillator model is about zero at the equilibrium state, i.e., $h_E+h_W=0$, one finds that $\alpha$ will be about half of $r$. The parameter $b_0$, which is the high-end estimation of the thermocline tilt and is in balance with the zonal wind stress produced by the SST anomaly, is chosen to give 50 m of east–west thermocline depth difference per $1^o$C of the SST anomaly. Thus, this model is nondimensionalized in a similar way as  the recharge paradigm, i.e., by scales of $[h] = 150$ m, $[T] = 7.5^o$C, $[u] = 1.5$ m/s, and $[t] = 2$ months for anomalous thermocline depth, SST, mixed-layer zonal current, and the time variables, respectively. Accordingly, parameters $c$, $r$, $\alpha_1$  and $\alpha_2$ are scaled by $1/[t]$, and parameters $\gamma$ and $b_0$ by $[h][t]/[T]$ and $[T]/[h]$. Their nondimensional values are $c = 1$, $\gamma = 0.75$, $\sigma = 0.6$, $r = 0.25$, $\alpha_1 = 0.0625$, $\alpha_2 = 0.125$ and $b_0 = 2.5$, which all correspond to those used in the recharge paradigm.

In the model, the relative coupling coefficient $\mu$ is $0.5$, which is smaller than the critical value (i.e., $0.7$ for purely oscillating). Such a choice implies that all the eigenvalues of this four-dimensional model have negative real part, representing the negative growth rates of the solution. It also allows the model to have a pair of conjugate solutions, i.e., damped oscillating solutions, which mimic the ENSO cycles. These features facilitate the stochastic excitation nature of the ENSO events by the random wind bursts. It has been shown that this conceptual model can depict the different variations between the CP and EP well \cite{fang2018three}. Specifically, with an increasing magnitude of the zonal advective feedback over the CP, i.e., imitating the situation for CP ENSO, the period of the system and SST magnitude over the CP and EP both decrease. Note that the decreasing amplitude is more intense over the EP, indicating an enlargement of the SST differences between the CP and EP. These results are all consistent with the observational characteristics of the CP El Ni\~no.

The model \eqref{Model_Deterministic} succeeds in describing the basic two-regime dynamical behavior of the ENSO for the EP and CP events. Yet, due to the deterministic nature, it cannot reproduce the observed irregularity of ENSO in amplitude and phase as well as the regime switching behavior and the non-Gaussian PDFs. Therefore, to simulate the realistic ENSO complexity,  additional processes in the intraseasonal and decadal timescales are further developed and coupled to the model \eqref{Model_Deterministic}.

\subsection*{The intraseasonal model for the random wind bursts}
The intraseasonal variability accounts for several important ENSO triggers, such as the WWBs, the EWBs, as well as the convective envelope of the MJO, which serve as the random input for the large-scale ENSO dynamics \cite{vecchi2006reassessing,chen2015strong,puy2016modulation,hu2016exceptionally}. The intraseasonal component here is modeled by a simple stochastic differential equation \eqref{Model_Stochastic_tau}. One important feature of \eqref{Model_Stochastic_tau} is that the noise coefficient $\sigma_\tau$ is state-dependent, i.e., a multiplicative noise. Here we assume it is positively correlated with $T_C$ since according to observations most of the wind bursts are active in the central-west Pacific \cite{tziperman2007quantifying, hendon2007seasonal, puy2016modulation}. The stochastic process \eqref{Model_Stochastic_tau} can generate both the WWBs and the EWBs, corresponding to $\tau$ (wind burst amplitude in the model) with positive and negative values, respectively. Notably, the state-dependent noise used here is very different from the previous two-region conceptual models, where the noise dependence was on $T_E$\cite{levine2010noise, perez2005comparison} due to the lack of the state variable $T_C$ in those models.

\begin{figure}[h]
\centering
\hspace*{-2cm}\includegraphics[width=18cm,angle=0]{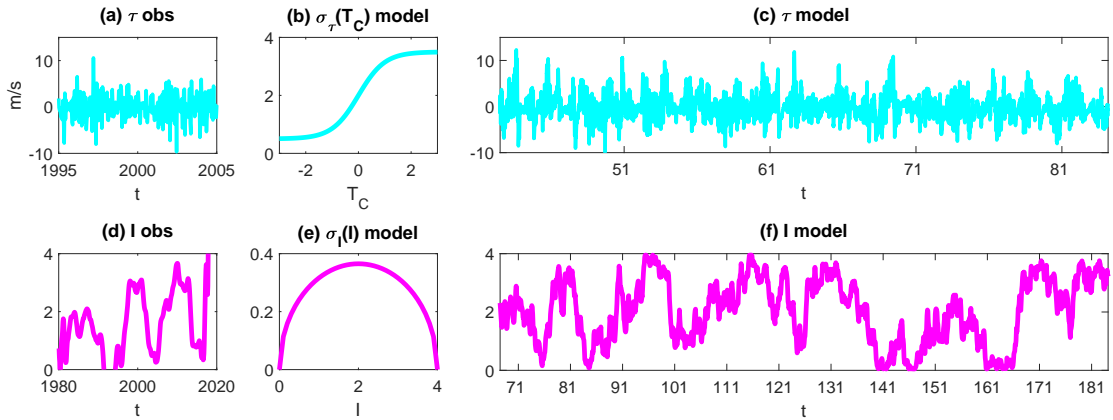}
\caption{Comparison between the observations and the model simulations for the intraseasonal wind burst variable $\tau$ and the decadal variable $I$. Panel (a): the observational wind burst time series. Panel (b): the multiplicative noise $\sigma_\tau$ in the stochastic process \eqref{Model_Stochastic_tau} for the intraseasonal variable $\tau$. Panel (c): the model simulation of $\tau$. Panels (d)-(f) are similar to Panels (a)-(c) but for the decadal variable $I$. In panel (d), the observational $I$ is smoothed by a 5-year window and then multiplied by a constant to make the final standard deviation be the same as the original one.}
\label{Variable_I}
\end{figure}

The intraseasonal component here is modeled by a simple stochastic differential equation \eqref{Model_Stochastic_tau}, which accounts for its intermittent and unpredictable nature at interannual timescale. The variable $\tau$ is the wind burst amplitude with a unit of $[\tau]=5$ m/s. The damping parameter $d_\tau=2$ representing a time scale of 1 month of the wind envelope but each individual wind is random in the daily time scale. The explicit expression of the noise coefficient is

\begin{equation}\label{Intraseasonal_Noise_Coeff}
	\sigma_\tau(T_C) = 0.9 [\tanh(7.5T_C) + 1],
\end{equation}
which clearly indicates a positive correlation between $T_C$ and $\sigma_\tau$. The reason for adopting a hyperbolic tangent function is to prevent the unbounded growth of $\sigma_\tau$ when the absolute value of $T_C$ becomes large. Note that $T_C$ here is the non-dimensional value. The profile of $\sigma_\tau(T_C)$ and a random realization of the simulated $\tau$ can be found in Figures \ref{Variable_I}b and \ref{Variable_I}c, respectively. The latter bears a high resemblance with the observations (Figure \ref{Variable_I}a).

\subsection*{The decadal model for the Walker circulation}
Several detailed El Ni\~no-type classification methods have been utilized to show that since 1870 the EP and CP events were alternatively prevalent every ten or twenty years \cite{yu2013identifying}. For example, the EP episodes were the dominant ones in the 1980s while the CP El Ni\~no events occurred more frequently since 2000 \cite{chen2015strong}. These findings indicate that the decadal variability plays an important role in driving the switching between the CP- and EP-dominant regimes. Thus, it is necessary to incorporate the decadal effect into the coupled ENSO model. The decadal model \eqref{Model_Stochastic_I} proposed here is a simple stochastic process, which aims at describing the large-scale behavior, including the characteristic time scale and the amplitude. For simplicity, the decadal variable $I$ is assumed to have no explicit dependence on the variables in the faster time scales, i.e., intraseasonal and interannual, but those ingredients are nevertheless effectively parameterized in the stochastic noise.  The damping parameter $\lambda=2/60$ is taken such that the decorrelation time $1/\lambda$ is about half of a decade. Next, despite that the observational data allows us to determine the range of $I$ being between $I=0$ and $I=4$, the $40$-year observational data is too short to provide unbiased information about the PDF of the decadal variability. Note that since $I$ is a slow-varying variable and it is bounded below by $I=0$, i.e., SST in the WP is warmer than the CP on the decadal timescale, it is unreasonable to assume a Gaussian distribution. Here, we adopt the uniform distribution function of $I$. This is based on the fact that the uniform distribution is the maximum entropy solution for a function in the finite interval without additional information \cite{kapur1992entropy}. Numerical tests have shown that replacing the uniform distribution by other empirically determined PDFs of $I$, such as a truncated Gaussian or a truncated bimodal distribution, only has a minor impact on the SST statistics provided that the decorrelation time of $I$ lies in the decadal time scale and the probability of $I$ at each point within the interval $[0,4]$ is non-vanishing. The resulting $\sigma_I(I)$ associated with the uniform distribution of $p(I)$ is included in Figure \ref{Variable_I}e. Figure \ref{Variable_I}f also shows a random realization of the time series $I$, which clearly indicates a stochastic regime switching behavior in the decadal time scale.

The stochastic decadal model is shown in \eqref{Model_Stochastic_I}, where $I$ is a surrogate of the decadal variation of the Walker circulation, and also the zonal SST difference between the WP and CP regions that directly determines the strength of the zonal advective feedback. In other words, $\sigma$ in \eqref{Model_Deterministic_T_C} can be regarded to be proportional to $I$. Specifically, $\sigma = 0.2I$ is used here, suggesting that it could give an SST rate of change of $1.5^o$C over $2$ months per $0.375$ m/s (when $I = 4$, i.e., the CP ENSO regime) or $1.5$ m/s (when $I = 1$, i.e., the EP ENSO regime) of the zonal current anomaly over the central Pacific.

\subsection*{Details of the three-region multiscale stochastic model for the ENSO complexity in \eqref{Model_Stochastic}}
\subsubsection*{Determining the nonlinearity in the coupled model}
Note that one difference between the starting deterministic model \eqref{Model_Deterministic} and the coupled stochastic model \eqref{Model_Stochastic} is the collective damping rates. The single damping coefficient in the deterministic model \eqref{Model_Deterministic} is splitted into two distinct values $c_1$ and $c_2$ in the governing equations of $T_C$ and $T_E$, respectively. The damping parameter

\begin{equation}\label{c2_equation}
    c_2=1.4
\end{equation}
in the $T_E$ equation remains as a constant since the WWBs are the main contribution to the positive skewness and the one-sided fat tail of the Ni\~no3 PDF corresponding to a large kurtosis (Figure \ref{PDFs_Spectrums}c). On the other hand, the ocean zonal current plays a more important role in the CP. Thus, the WWBs are not the main mechanism for the non-Gaussian statistics of the SST in the CP region since otherwise the associated PDF would have a similar profile as that of $T_E$, which is however not the case for the PDF associated with the observational data. In fact, the PDF of the observed Ni\~no4 SST has a different skewness direction and it has no fat tail (Figure \ref{PDFs_Spectrums}d).

\begin{figure}[h]
\centering
\hspace{-0.5cm}\includegraphics[width=10cm,angle=0]{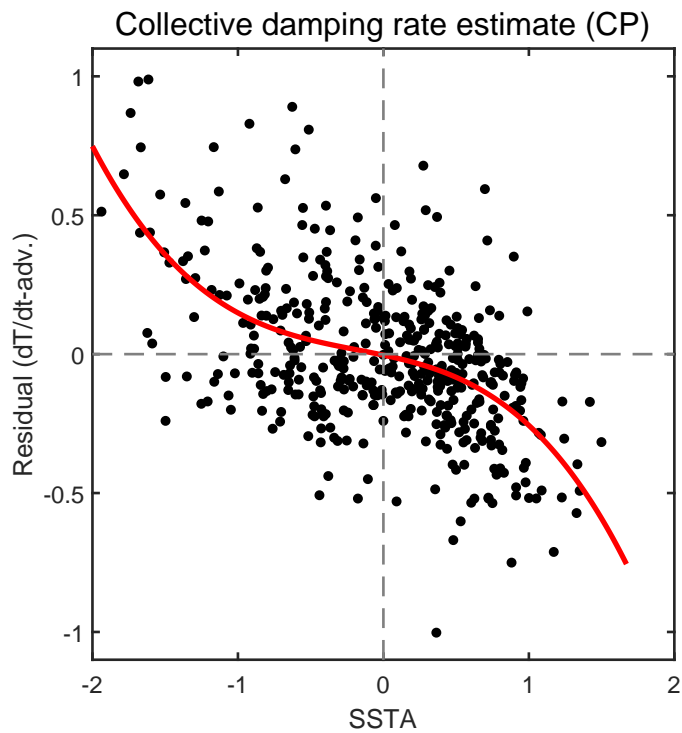}
\caption{The scatter plot: the relationship between the SST anomaly in the CP region, i.e., $T_C$, and the residual in the temperature equation \eqref{heat_budget}. Each black dot represents the value for one month. The black curve: the cubic polynomial fit of the scatter plot. The equation of the curve is $f(T_C) = -0.09{T_C}^3-0.05{T_C}^2-0.11{T_C}-0.01$, which passed the 95\% confidence from a Student's $t$ test. }
\label{c1_estimate}
\end{figure}

To understand the contributor of the nonlinear and non-Gaussian features in the CP region, a heat budget analysis of the mixed layer temperature is performed as follows:

\begin{equation}\label{heat_budget}
	\frac{\partial T_C^{\prime}}{\partial t} = -u^{\prime}\frac{\partial \bar{T}_C}{\partial x}-v^{\prime}\frac{\partial \bar{T}_C}{\partial y}-w^{\prime}\frac{\partial \bar{T}_C}{\partial z}-u^{\prime}\frac{\partial T_C^{\prime}}{\partial x}-v^{\prime}\frac{\partial T_C^{\prime}}{\partial y}-w^{\prime}\frac{\partial T_C^{\prime}}{\partial z}-\bar{u}\frac{\partial T_C^{\prime}}{\partial x}-\bar{v}\frac{\partial T_C^{\prime}}{\partial y}-\bar{w}\frac{\partial T_C^{\prime}}{\partial z}+Res,
\end{equation}
where overbars and primes indicate monthly climatology and anomaly, respectively. The variables $u$, $v$ and $T_C$ indicate zonal current, meridional current and oceanic temperature averaged over the mixed layer (top $50$ m). The vertical velocity ($w$) is calculated at the bottom of the mixed layer and $Res$ is the residual term, which represents the collective damping \cite{zebiak1987model}. Figure \ref{c1_estimate} shows a scatter plot of this residual term as a function of $T_C$ for the data at each month during the satellite observational era, which exhibits a clear nonlinear dependence. In fact, a cubic polynomial fitting curve in terms of $T_C$, also shown in the figure, well represents the relationship. It implies that $c_1$ is small for the small local SST anomalies and becomes larger for the large SST anomalies. This physically corresponds to the fact that larger damping will emerge (e.g., strong wind and precipitation) when the underlying SST is too warm \cite{xie2020new}. Thus, in the coupled model \eqref{Model_Stochastic}, $c_1$ is depicted by a simple nonlinear equation,

	\begin{equation}\label{c1_equation}
			c_1(T_C) = 25\left(T_C+\frac{0.75}{7.5}\right)^2+0.9.
	\end{equation}
Note that $c_1(T_C)$ is not centered at $T_C=0$, which is consistent with the observational data in Figure \ref{c1_estimate}.
Recall that the dimension of $T_C$ is $[T]=7.5^o$C. Therefore, $c_1$ is centered at $-0.75^o$C. This non-zero center explains the asymmetry of the SST in the CP region, which leads to a negative skewness since the damping with positive $T_C$ is overall stronger than that with the negative one. In addition, this cubic damping prevents strong $T_C$ in both positive and negative sides, which facilitates a kurtosis of the Ni\~no4 SST distribution that is smaller than $3$. This is consistent with the observations and it is also distinguished from the large kurtosis in the Ni\~no3 SST.

Next, the strength of the zonal advective feedback is depicted by $\sigma$, with a modulation by the decadal variation of the Walker circulation. In fact, the zonal advective feedback is

        \begin{equation}\label{zonal_advective_feedback}
			\frac{\partial T_C^{\prime}}{\partial t} =-u^{\prime}\frac{\partial \bar{T}_C}{\partial x},
		\end{equation}
where $\partial \bar{T}_C$ is the background zonal SST difference between the WP and CP, which shows to directly control $\sigma$ with a linear relationship and can be depicted by the decadal model  \eqref{Model_Stochastic_I} of variable $I$. Specifically, if $\partial \bar{T}_C$ is $3^o$C (i.e., $I = 3$), as provided in the standard setting, it could give an SST rate of change of $1.5^o$C over nearly $2$ months per $0.5$ m/s of the zonal current anomaly over the central Pacific, since the distance between the WP and CP is fixed ($50^o$ longitude). Under this situation, $\sigma$ will be 0.6 according to the nondimentional values. As a result, a simple relationship between $\sigma$ and $I$ can be derived, that is, $\sigma = 0.2I$.

Finally, the correction coefficient $C_u = 0.03$ is adopted in the coupled model.


\subsubsection*{Coupling coefficients between the interannual variables and the wind bursts}
In the model \eqref{Model_Stochastic}, the coupling coefficients $\beta_u$, $\beta_h$, $\beta_C$ and $\beta_E$ are determined systematically via the eigenmodes of the model \eqref{Model_Deterministic}. Specifically, since model \eqref{Model_Deterministic} is characterized by a pair of the damped oscillating modes, any external forcing that is imposed on this characteristic direction can be distributed to each of its four components $u$, $h_W$, $T_C$ and $T_E$ by multiplying the corresponding component of the eigenvector. A direct calculation shows that

\begin{equation*}
    (\beta_u, \beta_h, \beta_C, \beta_E)^T = (-0.2,-0.4,0.8,1.0)^T\beta_E,
\end{equation*}
which implies the response of $T_C$ is positively correlated to that of $T_E$  due to the wind burst forcing while the changes in $h_W$ and $u$ are anti-correlated with the SST response. These are all consistent with physics. Here the coefficient $\beta_E$ (before including the seasonal phase locking effect) takes the value

\begin{equation*}
    \beta_E = 0.15 \left(2-\frac{I}{5}\right).
\end{equation*}
It implies that $\beta_E$ increases as the decadal variable $I$ decreases. In other words, the wind bursts becomes stronger when $I$ favors the EP-dominant regime.

Finally, the values of the white noise strengths are

\begin{equation*}
    \sigma_u = 0.04\qquad \sigma_h = 0.02 \qquad\sigma_C = 0.04 \qquad\mbox{and}\qquad \sigma_E = 0.
\end{equation*}
Here, the uncertainties in $u$ and $T_C$ are the largest since their actual intrinsic processes are more complicated than the simple structures used here \cite{mccreary1981linear,zebiak1987model}. On the other hand, no additional noise is imposed to the equation of $T_E$ because $T_E$ is largely modulated by the random wind bursts.

\subsubsection*{Seasonal phase locking}
Seasonal phase locking is one of the remarkable features of ENSO, which manifests in the tendency of ENSO events to peak during boreal winter and is mainly related to the pronounced seasonal cycle of mean state \cite{tziperman1997mechanisms,stein2014enso}. Specifically, in the central-eastern Pacific, the climatological SST cools in boreal fall and warms in spring as a result of the seasonal motion of the ITCZ, which also modulates the strength of the upwelling and horizontal advection processes to influence the evolution of the SST anomalies \cite{mitchell1992annual}. Since the cool (warm) SSTs tend to coincide with decreased (increased) convective activity and upper cloud cover, a season-dependent damping term, which represents the cloud radiative feedback, can account for this seasonal variation in a simple fashion \cite{thual2017seasonal}. Besides, the increased wind burst activity in winter as a direct response to the increased atmospheric intraseasonal variability such as the MJO is the main constitution of the seasonal cycle in the western Pacific \cite{hendon2007seasonal, seiki2007westerly}. As a result, the seasonal cycle effects can be incorporated into the parameters $\sigma_\tau(T_C)$ in \eqref{Intraseasonal_Noise_Coeff}, $c_1(T_C)$ in \eqref{c1_equation} and $c_2$ in \eqref{c2_equation},

\begin{equation}\label{Seasonal_Coefficients}
\begin{split}
	\sigma_\tau(T_C,t) &= 0.9 [\tanh(7.5T_C) + 1]\left[1+0.3\cos\left(\frac{2\pi}{6}t+\frac{2\pi}{6}\right)\right],\\
    c_1(T_C,t) &= \left[25\left(T_C+\frac{0.75}{7.5}\right)^2+0.9\right]\left[1+0.3\sin\left(\frac{2\pi}{6}t-\frac{2\pi}{6}\right)\right],\\
    c_2(t) &= 1.4\left[1+0.2\sin\left(\frac{2\pi}{6}t+\frac{2\pi}{6}\right)+0.15\sin\left(\frac{2\pi}{3}t+\frac{2\pi}{6}\right)\right].
\end{split}
\end{equation}
Recall that the time unit $[t]$ is $2$ months. Therefore, $t$ going from $0$ to $6$ completes one year, where $t=0$ corresponds to January. Note from \eqref{Seasonal_Coefficients} that the strength of the wind bursts peaks in boreal winter, which is consistent with observations for the WWBs and the MJO \cite{hendon2007seasonal, zhang2005madden}. The second sinusoidal function in the collective damping $c_2$ represents a semiannual contribution to the seasonally modulated variance, as was suggested in a previous work \cite{stein2014enso}, which does not directly link with a semiannual cycle of the SST itself.

\bibliography{references}



\section*{Acknowledgements}

The research of N.C. is partially funded by the Office of VCRGE at UW-Madison. The research of X.F. is supported by the Ministry of Science and Technology of the People's Republic of China (Grant No. 2020YFA0608802) and the National Natural Science Foundation of China (Grant No. 41805045).

\section*{Author contributions statement}

N.C. and X.F. designed the project. N.C., X.F. and J.Y. performed the research. N.C. and X.F. wrote the paper. 

\section*{Additional information}
\textbf{Data availability} The monthly ocean temperature and current data were downloaded from the National Centers for Environmental Prediction Global Ocean Data Assimilation System  (https://www.esrl.noaa.gov/psd/data/gridded/data.godas.html). The daily zonal wind data at 850 hPa were downloaded from the NCEP–NCAR reanalysis \\ (https://psl.noaa.gov/data/gridded/data.ncep.reanalysis.html).

\noindent \textbf{Competing interests} The authors declare no competing interests.

\clearpage

\clearpage

\begin{table}[h]
\centering
\begin{tabular}{|l|c|l|c|}
  \hline
  $[h]$ & $150$ m & $[T]$ & $7.5 ^o$C\\
  $[u]$ & $1.5$ m/s & $[t]$ & $2$ months\\
  $[\tau]$ & $5$ m/s & $d_\tau$ & $2$\\
  $\gamma$ & $0.75$ & $r$ & $0.25$ \\
  $\alpha_1$ & $0.0625$ & $\alpha_2$ & $0.125$ \\
  $b_0$ & $2.5$ & $\mu$ & $0.5$ \\
  $\sigma$ & $0.2I$ & $\lambda$ & $0.1$ \\
  $p(I)$ & $0.25$ in $I\in(0,4)$& $\sigma_I(I)$ & Fig. \ref{Variable_I}e\\
  $\beta_E$ & 0.15(2-0.2I) & $\beta_u$ & $-0.2\beta_E$ \\
  $\beta_h$ & $-0.4\beta_E$ & $\beta_C$ & $0.8\beta_E$ \\
  $\sigma_u$ & $0.04$ & $\sigma_h$ & $0.02$ \\
  $\sigma_C$ & $0.04$ & $\sigma_E$ & $0$ \\
  \hline
  $\sigma_\tau(T_C,t)$ & \multicolumn{3}{l|}{$0.9 [\tanh(7.5T_C) + 1]\left[1+0.3\cos\left(\frac{2\pi}{6}t+\frac{2\pi}{6}\right)\right]$} \\
  $c_1(T_C,t)$& \multicolumn{3}{l|}{$\left[25\left(T_C+\frac{0.75}{7.5}\right)^2+0.9\right]\left[1+0.3\sin\left(\frac{2\pi}{6}t-\frac{2\pi}{6}\right)\right]$}\\
  $c_2(t)$& \multicolumn{3}{l|}{$1.4\left[1+0.2\sin\left(\frac{2\pi}{6}t+\frac{2\pi}{6}\right)+0.15\sin\left(\frac{2\pi}{3}t+\frac{2\pi}{6}\right)\right]$}\\
  \hline
\end{tabular}\caption{Summary of the non-dimensional units and the model parameters.}\label{Table_Parameters}
\end{table}

\begin{table}[h]
\centering
\begin{tabular}{@{}rrrrrrrr@{}}
\toprule
 & El Ni\~no & EP & CP & Extreme & Multi-year & La Ni\~na & Multi-year \\ \midrule
 Observation (1950-2020)   & 24(34\%)  & 14(20\%)  & 10(14\%)   & 4(6\%)   & 5(7\%)   & 24(34\%)  & 8(11\%)   \\
Standard run (2000 years) & 660(33\%) & 398(20\%) & 262(13\%) & 125(6\%) & 100(5\%) & 852(43\%) & 209(10\%) \\
$I\equiv0$ (2000 years)                       & 646(32\%) & 453(23\%) & 193(9\%) & 260(13\%) & 78(4\%)  & 861(43\%) & 200(10\%) \\
$I\equiv4$ (2000 years)                       & 769(38\%) & 441(22\%) & 328(16\%) & 34(2\%)  & 175(9\%) & 782(39\%) & 196(11\%) \\
 \bottomrule
\end{tabular}\caption{Comparison of the model results for different situations with the observation on the ENSO complexity. Shown are the numbers of different ENSO events and their percentages in the whole periods, i.e., 2000 years and 71 years are respectively for the model and observation.}\label{Table_complexity}
\end{table}

\end{document}